\documentclass[twocolumn,amsmath,amssymb,floatfix,prl]{revtex4}

\usepackage{graphicx}           
\usepackage{float}
\usepackage{amsmath}
\SetSymbolFont{largesymbols}{normal}{OMX}{iwona}{m}{n}

\def\omr{\Omega_{\rm{R}}}
\def\kr{\ k_{\rm{L}}}
\def\Er{\ E_{\rm{L}}}

\def\delp{\delta^{\prime}}

\DeclareMathAlphabet\mathbfcal{OMS}{cmsy}{b}{n}

\def\nm{{\ {\rm nm}}}                       
\def\um{{\ \mu{\rm m}}}                 

\def\kHz{{\ {\rm kHz}}}                     
\def\MHz{{\ {\rm MHz}}}                     

\def\ms{{\ {\rm ms}}}                       

\def\Er{{{E_L}}}                            
\def\kr{{{k_L}}}                            
\def\Rb87{^{87}\rm{Rb}}                 
\def\Li6{^{6}\rm{Li}}                   

\def\ex{{\mathbf e}_x}                            
\def\ey{{\mathbf e}_y}                            
\def\ez{{\mathbf e}_z}                            
\DeclareMathAlphabet\mathbfcal{OMS}{cmsy}{b}{n}

\begin{document}

\title{Vortex Nucleation in a Bose-Einstein Condensate: From the Inside Out}

\author{R.~M.~Price}
\affiliation{Joint Quantum Institute, University of Maryland and 
	National Institute of Standards and Technology, College
	Park, Maryland, 20742, USA}

\author{D.~Trypogeorgos}
\affiliation{Joint Quantum Institute, University of Maryland and 
National Institute of Standards and Technology, College
Park, Maryland, 20742, USA}

\author{D.~L.~Campbell}
\affiliation{Joint Quantum Institute, University of Maryland and 
National Institute of Standards and Technology, College
Park, Maryland, 20742, USA}

\author{A.~Putra}
\affiliation{Joint Quantum Institute, University of Maryland and 
National Institute of Standards and Technology, College
Park, Maryland, 20742, USA}

\author{A. Vald\'{e}s-Curiel}
\affiliation{Joint Quantum Institute, University of Maryland and 
	National Institute of Standards and Technology, College
	Park, Maryland, 20742, USA}

\author{I.~B.~Spielman}
\affiliation{Joint Quantum Institute, University of Maryland and 
	National Institute of Standards and Technology, College
	Park, Maryland, 20742, USA}
\date{\today}

\begin{abstract}
	We observed a new mechanism for vortex nucleation in Bose-Einstein condensates (BECs) subject to synthetic magnetic fields. We made use of a strong synthetic magnetic field initially localized between a pair of merging BECs to rapidly create vortices in the bulk of the merged condensate. Unlike previous implementations and in agreement with our Gross-Pitaevskii equation simulations, our dynamical process rapidly injects vortices into our system's bulk, and with initial number in excess of the system's equilibrium vortex number.
\end{abstract}


\maketitle
\begin{figure*}[bt]
	\begin{center}
		\includegraphics{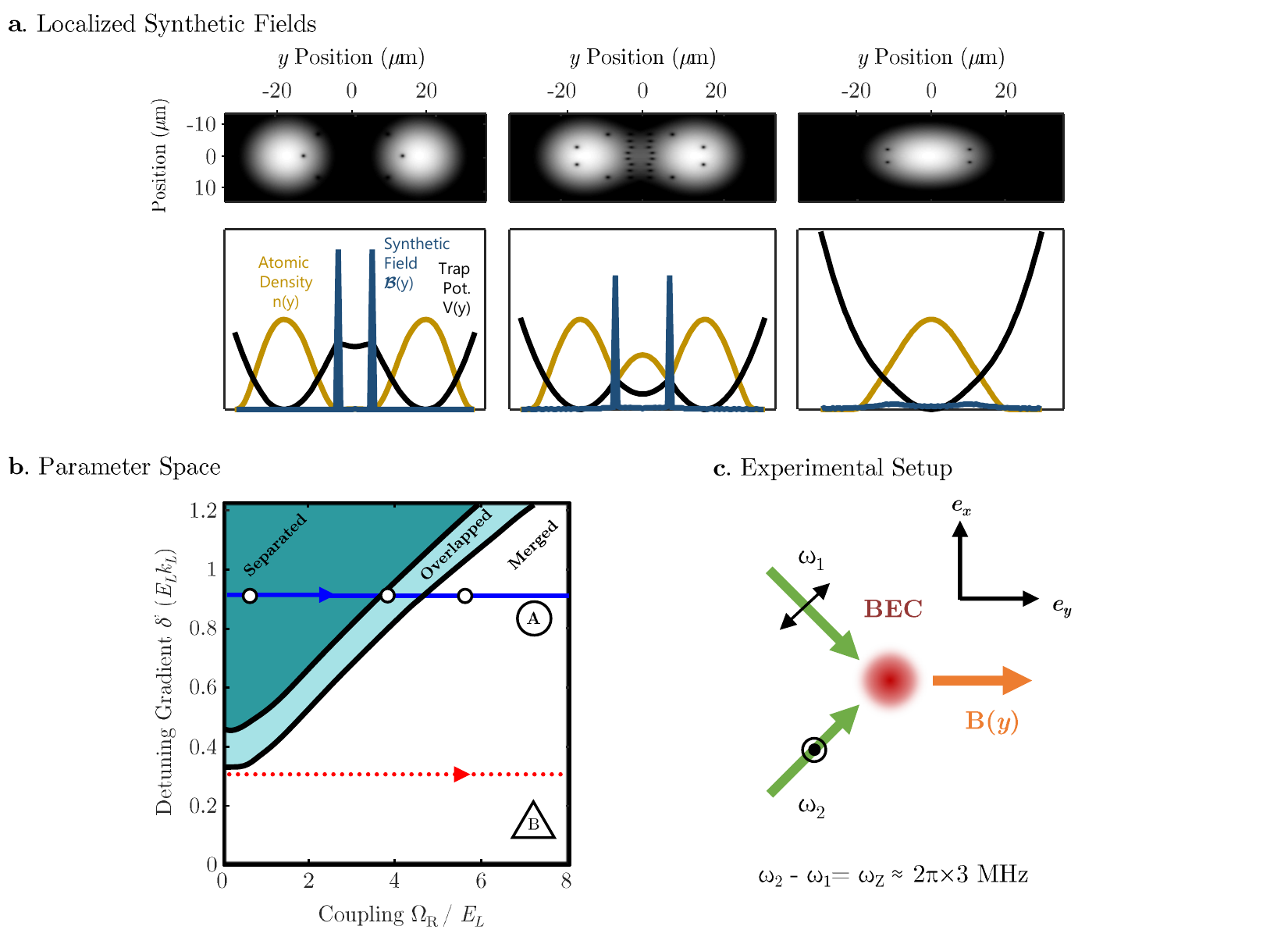}
		\caption
		{
			{\bf a)} Top: GPE-computed 2D density distributions $n(x,y)$. Columns correspond to Raman coupling strengths $\hbar\omr / E_L = 0.5,\,3.5,\,5.5 $ respectively, all with the detuning gradient $\hbar\delp = 0.92 \, E_{\rm{L}} k_{\rm{L}}$. Bottom: Computed cross-sectional cuts of the total potential $V(y)$ (black), effective magnetic field ${\mathcal B}$(y) (blue) and atomic density $n(y)$ computed in the Thomas-Fermi approximation (gold) using the same parameters as above.		
			{\bf b)} Parameter space for $\delp$ and $\omr$ for our system configuration. The parameters can yield regions where the two condensates are separated (regime I, teal), overlapped (regime II, light blue), or merged (regime III, white). Indicators along trajectory a represent parameters depicted in panel a.
			{\bf c)}  $\Rb87$ BECs in a linearly varying magnetic field ${\bf B} = (B_0 + B' y)\ey$ illuminated by a pair of cross-polarized Raman laser beams at $\omega_{R}$ propagating along $\ex \pm \ey$.
		}
		\label{fig:Figure1}
	\end{center}
\end{figure*}

Degenerate ultracold atomic gases are versatile quantum fluids which can have behavior analogous to those present in other quantum systems, ranging from superconductors \cite{bloch2012quantum,greiner2008condensed} to neutron stars \cite{zwierlein2005vortices}. Quantized vortices are a common element present in superconductors \cite{trauble1968flux}, superfluids \cite{packard1972observations}, dilute atomic Bose-Einstein condensates (BECs) \cite{Matthews1999}, and any other system where the single-valuedness of the wavefunction demands quantized circulation. The high degree of control provided by atomic systems makes them unique for studying vortex physics. Since the first creation of vortices in BECs \cite{Matthews1999}, many experiments have investigated vortex formation and dynamics. While a few vortices were created in atomic BECs by directly engineering an appropriate atomic wavefunction \cite{Matthews1999,Leanhardt2002}, large ensembles are typically only present in rapidly rotating systems \cite{Madison2000,abo2001observation,Engels2003,andersen2006quantized}. For rotating BECs, vortices nucleate on the system's periphery, then migrate into the bulk and ultimately form a vortex lattice, typically a slow process. Synthetic magnetic fields can offer a different means to inject vortices in BECs, but in initial experiments \cite{lin2009synthetic} the mechanism for vortex formation was the same as in rotating systems. Here we extended those studies and demonstrate a new mechanism for vortex nucleation in which vortices are rapidly formed within the system's interior. 

In quantum fluids, quantized vortices result from the requirement that the wavefunction be both continuous and single-valued.  Along any closed path the wavefunction's phase can only advance by integer multiples of $2\pi$.   For example, the phase advances by $2\pi$ for paths encircling a singly `charged' vortex, giving $\hbar$ angular momentum per particle; to maintain its continuity, the wavefunction vanishes at the vortex center. Early rotation experiments exploited the equivalence of the Lorentz force and the Coriolis force in rotating systems to generate uniform artificial magnetic fields ${\mathbfcal B} = {\mathcal B}\ez$. In those experiments, the rate at which vortices enter the system and their number in equilibrium are dependent on the rate of rotation and the condensate size \cite{dalfovo1999,chevy2002formation,stock2005bose}. For BECs with repulsive interactions, it is energetically favorable for vortices to form at the system’s edge, where the low atomic density facilitates vortex nucleation. These vortices then migrate toward the center of the condensate, where they can ultimately equilibrate into a vortex lattice. Even in cases where the effective magnetic field is not uniform across the condensate, the same mechanism of vortex nucleation applies \cite{murray2009vortex}. The vortex density across the condensate will be correlated with the geometry of the effective magnetic field, with vortices congregating in high-field regions.

Here we created an inhomogeneous laser-induced artificial magnetic field  \cite{spielman2009raman} initially maximized in the space between a pair of spatially separated BECs. The atomic density in the localized high-field region was small but non-zero, allowing the ready formation of precursor vortices in regions of negligible atomic density \cite{wen2010hidden}. We then gradually expanded the region of high-field while merging the BECs, culminating with a single BEC with a nominally uniform field, incorporating the precursor vortices into the BEC’s center during the merging process. 

We explored regimes where both uniform and non-uniform effective fields can exist, as shown in Fig.~\ref{fig:Figure1}a. We started with $\Rb87$ condensates in the $f = 1$ hyperfine ground state with $N\approx 4\times 10^5$ atoms in a $1064 \nm$ crossed optical dipole trap giving potential $V_{\rm{trap}}$ with frequencies $[\omega_x,\omega_y,\omega_z]/2\pi=[42(2),43(2),133(3)]\ {\rm Hz}$ \footnote{All uncertainties herein reflect the uncorrelated combination of single-sigma statistical and systematic uncertainties.}. 
We subjected the BECs to a linearly varying magnetic field ${\bf B} = (B_0 + B' y)\ey$ which gave a position-dependent Zeeman splitting $\hbar\omega_{\rm Z}(y) =  g_F\mu_{\rm B} |{\bf B}(y)|$ between the three $m_F$ states, with $g\mu_{\rm{B}}B_{0}/h = \omega_{\rm{Z}} / 2\pi \approx 3\MHz$. To create the synthetic magnetic field \cite{lin2009synthetic}, we illuminated the BEC with a pair of intersecting cross-polarized Raman laser beams of wavelength $\lambda_{\rm{R}} = 790.024(5)\nm$ propagating along $\ex \pm \ey$ with two-photon Raman coupling strength $\omr$, and differing in frequency by $\omega_{Z}$. The frequency difference sets the position dependent detuning $\delta(y) = g\mu_{\rm{B}} B^{\prime}y = \delp y$ with the laser frequency defining the single-photon recoil energy $E_{L} = \hbar^{2}k_{L}^2 / 2m \approx \hbar \times 1.8 \kHz$ and momentum $k_{L} = \sqrt{2}\pi / \lambda_{\rm{R}}$, where $m$ is the atomic mass. This configuration produces an artificial magnetic field ${\mathbfcal B} = {\mathcal B}\ez$ with strength set by $\omr$ and $\delp$ resulting from an artificial vector potential ${\mathbfcal A}$. Along with a scalar potential $\phi$, this gives an effective Hamiltonian for our BEC:

\begin{equation}
\begin{split}
H = \frac{\hbar^2}{2m^{*}}  &\left(k_x - \frac{{\mathbfcal A}(y;\omr, \delp)}{\hbar}   \right)^{2} + \frac{\hbar^2}{2m}  \left(k_{y}^2 + k_{z}^2\right)\\
&+ \phi(y; \omr, \delp) + V_{\rm{trap}}
\end{split}
\end{equation}

Where $m^{*}$ is an effective mass and both ${\mathbfcal A}$ and $\phi$ depend on $y$, and the $y$-dependence of ${\mathbfcal A}$ gives rise to ${\mathbfcal B}$. 
 
\begin{figure}[bth!]
	\begin{center}
		\includegraphics{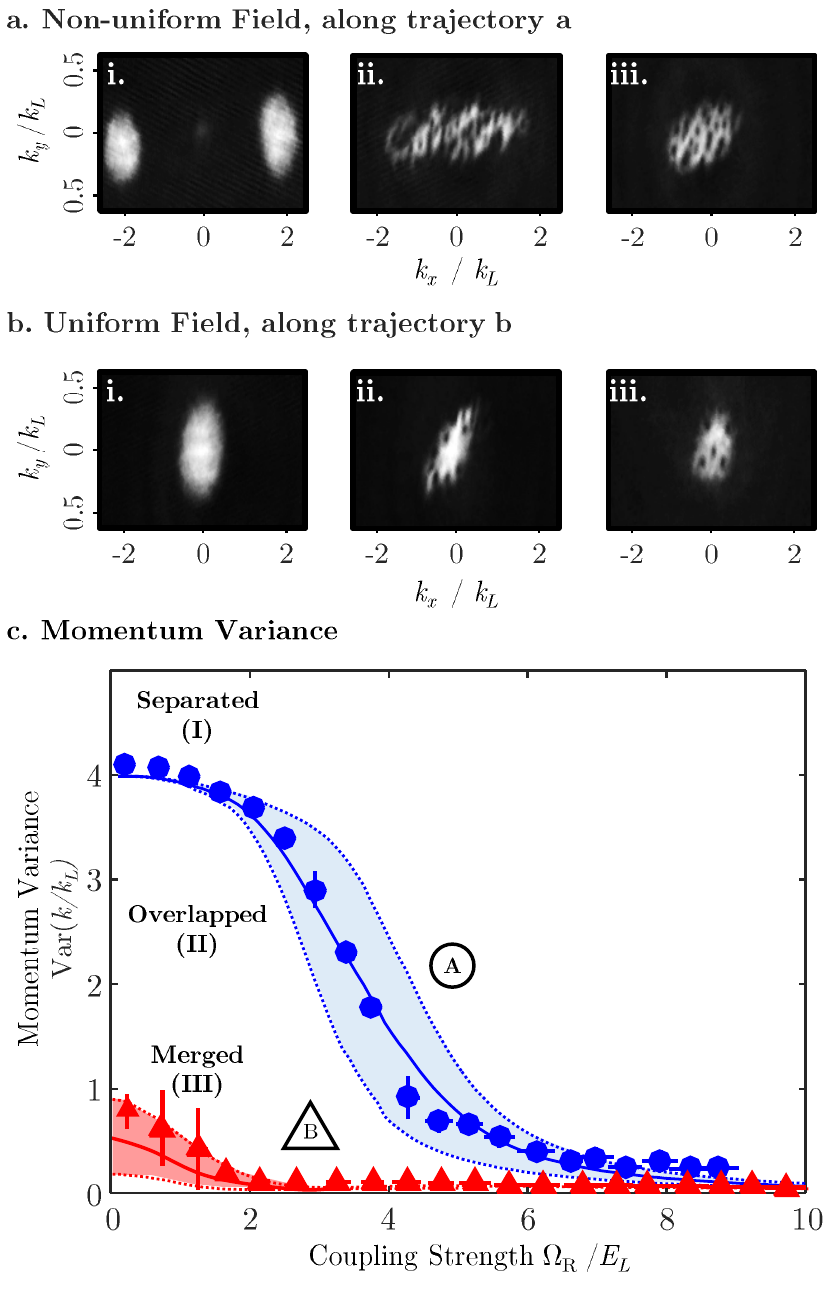}
		\caption{
			{\bf a)} TOF expanded images of BECs along trajectory a (non-uniform field), with $\delp = 0.92(4) \,E_{\rm{L}} k_{\rm{L}}$ at $\omr / \Er = 0.25, 3.81, 9$ for i, ii, and iii respectively.		
			{\bf b)} TOF expanded images of BECs along trajectory b (uniform field), with $\delp = 0.3(4) \, E_{\rm{L}} k_{\rm{L}}$ at $\omr / \Er = 0.25, 2.5, 9$ for i, ii, and iii respectively.
			{\bf c)} Momentum variance measured as a function of $\omr$ at $\delp = 0.92 \, E_{\rm{L}} k_{\rm{L}}$ (blue circles) and at $\delp = 0.3(4)$ (red triangles). Lines represent the theoretical model while the shaded region represents the range of uncertainty given our system parameters.			
		}
		\label{fig:Figure2}
	\end{center}
\end{figure}

Figure \ref{fig:Figure1}a shows ${\mathcal B}(y)$, along with the potential $V(y)$, the sum of $\phi(y)$ and the overall harmonic confining potential, showing three qualitatively different regimes (I, II and III) that depend on $\omr$ and $\delp$. In the regime I, two BECs are separated by a potential barrier containing the artificial field (Fig.~\ref{fig:Figure1}a, left) that hardly impacts the individual BECs. In regime II, the energy barrier between the two potential wells falls below the BECs' chemical potential $\mu$, resulting in a non-negligible atom density in the barrier regime containing the strong synthetic field (Fig.~\ref{fig:Figure1}a, center). As predicted by previous simulations \cite{Radic2011}, within this region the strong magnetic field gives a cyclotron energy of $\sim 1 \, \Er$ that locally dominates all other energy scales, readily forming a linear chain of vortices \cite{williams2010observation}. In regime III, the barrier has vanished and the artificial field has expanded, resulting a single BEC subject to a nominally uniform field (Fig.~\ref{fig:Figure1}a, right), akin to rotation experiments. As shown in Fig.~\ref{fig:Figure1}b, for sufficiently large detuning gradient $\delp$, these three regimes can be accessed sequentially with increasing $\omr$. Fixing $\delp$ and sweeping $\omr$ at a constant $\delp$ through these regimes drives a transition from non-uniform to nominally uniform field, releasing precursor vortices formed between the two BECs into the system's center. 

To study these regimes, we prepared our BEC with an equal fraction of atoms in the $m_F=\pm1$ states and linearly ramped on the detuning gradient from zero to a desired final value $\delta^{\prime}$ over $0.5 \:\rm{s}$, spatially separating the two spin components. We then waited $100 \ms$ for the magnetic field environment to equilibrate~\footnote{We believe this timescale is set by the damping of eddy currents in nearby conductors. Our gradient coils produced a small unwanted contribution to the bias field which we compensated for by adjusting the current in our bias coils, thereby keeping the bias at the system's center constant. Our time-of-flight data is highly sensitive to the trapping potential and detuning breaks the degeneracy between the potential wells: skewing the data.}, and then linearly ramped on the Raman coupling to $\omr$. 

The vortex core size is approximately the BEC healing length, $\xi = \hbar / \sqrt{2 m \mu} \approx 0.32(2) \um$, which is well below our system's $1.9 \um$ imaging resolution. Therefore we used time-of-flight (TOF) techniques to expand the cloud before absorption imaging, giving images approximating the momentum distribution. We initiated TOF by suddenly turning off the confining potentials, and in the first 2 ms of TOF we ramped $\omr$ to zero while simultaneously ramping the detuning $75 \:E_L / \hbar$ from resonance.  This process mapped the laser-dressed system into a single spin state, with a spatially uniform vector potential ${\mathbfcal A}_f$. The resulting spatially dependent change ${\mathbfcal A}(y) - {\mathbfcal A}_f$ imparted a position-dependent artificial electric field inducing an overall shearing motion \cite{leblanc2015gauge}. 

The BEC's momentum distribution is drastically different in each of the three parameter regimes in Fig.~\ref{fig:Figure1}a. This difference is well quantified by the variance of the momentum distribution ${\rm Var}(k)$.  In regime I, when there are two separated BECs (Fig.~\ref{fig:Figure2}a, left column) the momentum distribution is sharply peaked at $\pm 2 \,\kr$, giving a large ${\rm Var}(k)\approx 4 \,\kr^2$. In regime II, when these BECs are partially merged (Fig.~\ref{fig:Figure2}a, center column), the momentum distribution spans the full regime from $-2 \,\kr$ to $2 \,\kr$, reducing ${\rm Var}(k)$.  In regime III, with a single fully merged BEC (Fig.~\ref{fig:Figure2}a right column, and all of Fig. 1a), the momentum distribution is sharply peaked at $k=0$, minimizing ${\rm Var}(k)$.

For these studies the Raman coupling $\omr$ was ramped on at $\sim 10 \,\Er / \rm{s}$ rate chosen to be adiabatic with respect to the system's center of mass dynamics and then held constant for $150 \ms$, such that the momentum distribution equilibrates (this was not adiabatic with respect to the slower time scale for vortex dynamics~\cite{abo2001observation}). Figure~\ref{fig:Figure2}c shows the dependence of ${\rm Var}(k)$ on $\omr$ at $\delta^\prime= 0.92 \, E_{\rm{L}} k_{\rm{L}}$ (blue) containing all three qualitative regimes outlined previously. For $\omr \lesssim 3 \,\Er$, ${\rm Var}(k)$ decreased slowly, as expected for the separated well configuration. As $\omr$ increases, the scalar potential $\phi$ begins to weaken and the two separated BECs start to merge. This merging happens when $ 3 \,\Er \lesssim \omr \lesssim 5 \,\Er $ and is correlated with a rapid decline in ${\rm Var}(k)$. When $\omr \gtrsim 5\,\Er $, $\phi$ becomes weak in comparison to the trapping potential, and the system forms a single well potential, causing ${\rm Var}(k)$ to approach zero. In contrast for $\delp = 0.3 \, E_{\rm{L}} k_{\rm{L}}$ (red) along trajectory b, the variance is always small and the system remains in regime III for the entire sweep.

\begin{figure}[bt]
	\begin{center}
		\includegraphics{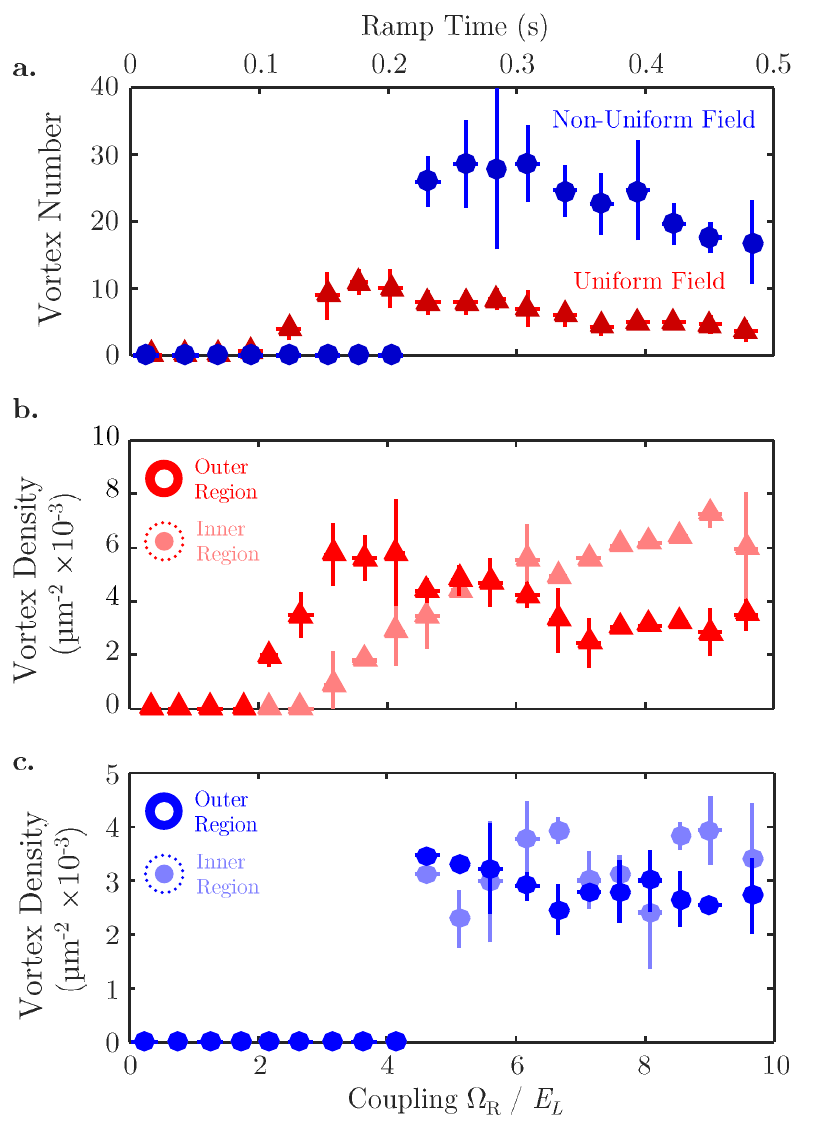}
	\end{center}
	\caption{		
		{\bf a)} Vortex number as a function of $\omr$ in the uniform (red triangles) and non-uniform (blue circles) effective field regimes (at $\delp = 0.92(4) \, E_{\rm{L}} k_{\rm{L}}, \: 0.3(4) \, E_{\rm{L}} k_{\rm{L}}$ respectively). The system was initially prepared at the experimental value of $\delp$ then $\omr$ was ramped up to a target value and the number of vortices in the BEC was measured. Uncertainty in vortex number results from difficulty in identifying low contrast vortices and the uncertainty in $\omr$ comes primarily from systematic uncertainties.
		{\bf b)} Vortex density in the outer (dark red) and inner halves (light red) of the BEC for the uniform field method regime.
		{\bf c)} Vortex density in the outer (dark blue) and inner halves (light blue) of the BEC for the non-uniform field method regime.		
	}
	\label{fig:Figure4}
\end{figure}

We now turn our focus to vortex nucleation.  As shown in Fig.~\ref{fig:Figure2}a-b, our TOF images can have many vortices -- a feature that distinguishes them from the true momentum distributions.  In TOF, interactions continue to play a role making vortices stable objects that persist and expand in TOF \footnote{Vortices are also preserved by the Fourier transform, so they generally will be there in true momentum distributions, but often with a qualitatively different appearance and position.}. To focus on vortices that were nucleated or injected into the BEC, we altered the preparation described previously, increasing the ramp rate of $10 \Er / \rm{s}$ to $\approx 20 \Er / \rm{s}$, again following the trajectories in Fig.~\ref{fig:Figure1}b.

Trajectory b only experienced regime III (the merged regime) while trajectory a crossed from regime I, through II, into III. The representative images in Fig.~\ref{fig:Figure2}b along trajectory b show a single condensate (panel i), which first nucleates vortices at its periphery (panel ii) before they finally enter into its bulk (panel iii). This behavior replicates that of previous rotating and synthetic field experiments~\cite{lin2009synthetic,Engels2003}. The images in Fig.~\ref{fig:Figure2}a shows contrasting behavior along trajectory a, in which two BECs (panel i) form numerous vortices as they merge (panel ii), that persist in large number in the merged BEC regime (panel iii). We quantify this behavior by locating and counting vortices in such images.

We developed a vortex identification algorithm (e.g. Ref.~\cite{rakonjac2015measuring}) that locates vortices in the central high-density region of the clouds. Our counting algorithm performs poorly for overlapped or low contrast vortices and this poor performance was particularly evident in distributions of partially merged condensates. For these cases, manual counting of vortices augmented the algorithm. Figure~\ref{fig:Figure4}a shows the result of such counting along trajectories a and b, confirming our prediction that vortices enter abruptly and in great number along trajectory a.

Along trajectory b where there was only a weak, uniform synthetic field, we observed a slow increase in the vortex number as $\omr$ was ramped up (Fig.~\ref{fig:Figure4}a, red). For larger $\omr$ where ${\mathcal B}$ begins to decrease \cite{lin2009synthetic,spielman2009raman}, the number of vortices also begins to fall \footnote{The reduced area of the BEC from larger spontaneous emission also contributed to the lower number of vortices.}. By comparison along trajectory a with a high-strength, non-uniform field, vortices appeared abruptly as $\omr$ was increased, before falling in number (Fig.~\ref{fig:Figure4}a, blue). The stark difference in rapid appearance of vortices, together with higher vortex number signify the different vortex nucleation mechanisms.

We distinguished these two potential mechanisms for vortex formation by studying the vortex density in the inner and outer regions of the system, delineated by half of the Thomas-Fermi radius. Since in conventional nucleation processes (trajectory b) vortices enter from the system's periphery (e.g. Fig.~\ref{fig:Figure2}b-ii), we expect the vortex density in the outer region to exceed that of the inner region while the vortices migrate inwards. In contrast along trajectory a, we expect vortices to be preformed in the system's center, quickly dispersing (Fig.~\ref{fig:Figure2}a-ii) across the BEC during the merging process.   

Along trajectory b (weak uniform field), the vortex density in the outer region of the BEC begins to increase before the vortex density in the inner region of the BEC (Fig.~\ref{fig:Figure4}b). The observation is that, similar to previous rotational experiments, the vortices are nucleated on the periphery of the BEC and evolve inward toward a lower energy state. For trajectory a (crossing a non-uniform field) the vortex density in both the inner and outer regions simultaneously increases approximately where our calculation predicts that the two spatial wells combine and spawn internal vortices (Fig.~\ref{fig:Figure4}c). This simultaneous increase is consistent with vortices preformed in the BEC's interior that then disperse across the condensate.

We experimentally demonstrated a novel nucleation mechanism that generates vortices from within the bulk of the system. This nucleation method rapidly generates a high number of vortices, which upon the full overlap of the two BECs, are quickly dispersed throughout the system. This experiment may be extended to rapidly generate vortices, much faster than typical equilibration times for a given trapping potential, before returning the system to a geometry without synthetic magnetic fields, allowing for the study of complex vortex nucleation hysteresis \cite{garcia2001vortex}, or superfluid turbulence \cite{henn2009emergence}. Furthermore this work could be extended with enhanced detection methods in order to observe the exotic equilibrium vortex structures predicted to form in regime II.

\begin{acknowledgements}
We appreciate constructive discussions with K. Helmerson.  This work was partially supported by the ARO's Atomtronics MURI, and by the AFOSR's Quantum Matter MURI, NIST, and the NSF through the PFC at the JQI. 
Our GPE calculations were performed with the open-source ``GPE Lab'' software \cite{Antoine2014}.
\end{acknowledgements}

\bibliography{VortexSOC}

\end{document}